\documentclass[12pt]{article}
\usepackage{graphicx}
\usepackage{float}
\newcommand{\y }{\'{\i}}
\newcommand{\be }{\begin{equation}}
\newcommand{\ee }{\end{equation}}
\begin{document}

\begin{center}
{\bf DECAY OF METASTABLE NONEQUILIBRIUM PHASES, ENHANCED REACTION RATE, AND 
DYNAMIC PHASE TRANSITION IN A MODEL OF CO OXIDATION WITH CO DESORPTION}\\
\vspace{15pt}
Erik Machado and Gloria M.~Buend\y a\\
{\it Physics Department, Universidad Sim\'on Bol\y var,\\
Apartado 89000, Caracas 1080, Venezuela}\\
\vspace{5pt}
Per Arne Rikvold\\
{\it Center for Materials Research and Technology,\\
School of Computational Science,  and Department of Physics, \\
Florida State University, Tallahassee, Florida 32306-4052, USA\\
and Department of Fundamental Sciences, 
Faculty of Integrated Human Studies, Kyoto University, Kyoto 606, Japan}\\
\vspace{5pt}
Robert M.~Ziff\\
{\it Department of Chemical Engineering and\\
Michigan Center for Theoretical Physics,\\
University of Michigan, Ann Arbor, MI 48109-2136, USA}
\end{center}

\vspace{15pt}
\begin{center}
ABSTRACT
\end{center}
%\vspace{5pt}
We present a computational study of the dynamic behavior of a
Ziff-Gulari-Barshad model of CO oxidation 
with CO desorption on a catalytic surface. 
Our results provide further evidence 
that below a critical desorption rate the model 
exhibits a non-equilibrium, first-order phase transition between low and high
CO coverage phases. Our kinetic Monte Carlo 
simulations indicate that the transition process between these phases
follows a decay  mechanism very similar to the one described by the 
classic Kolmogorov-Johnson-Mehl-Avrami
theory of phase transformation by nucleation and growth. 
We measure the lifetimes of the metastable phases on each side of 
the transition line and find that they
are strongly dependent on the direction of the transformation, 
i.e., from low to high coverage or 
vice versa. Inspired by this asymmetry, we introduce a square-wave periodic
external forcing, whose two parameters can be tuned to enhance the
catalytic activity. 
At CO desorption rates below the critical value, we find that 
this far-from-equilibrium system undergoes a dynamic phase transition
between a CO$_2$ productive phase and a nonproductive one. 
In the space of the parameters of the periodic external forcing, this 
nonequilibrium phase transition defines a line of critical points. The 
maximum enhancement rate for the CO$_2$ production rate occurs near 
this critical line.

%\newpage
\vspace{15pt}
\begin{center}
INTRODUCTION
\end{center}
%\vspace{5pt}
The study of phase transitions and critical phenomena in nonequilibrium 
statistical systems have recently attracted a great deal of attention due its
applications in many branches of physics, chemistry, biology, economics, and 
even sociology [1]. Within this field, surface reaction models have become
an archetype for studying out-of-equilibrium critical phenomena, 
and they have been intensely
analyzed with the purpose of designing more efficient catalytic processes [2]. 
The Ziff, Gulari, Barshad (ZGB) model [3] with desorption 
(the ZGB-k model) [4,5,6] describes kinetic aspects of the
gas-phase reaction CO+O $\rightarrow$ CO$_2$ on
a catalytic surface in terms of two parameters: the relative partial
pressure of CO, $y$, that represents the probability that the next
molecule arriving to the surface is a CO, and the desorption rate, $k$, 
which is related to 
the probability that an adsorbed CO molecule is desorbed without being oxidized. 
The overall reaction is assumed to occur according to the 
Langmuir-Hinshelwood mechanism,
\begin{eqnarray}
\mathrm{CO(g)}+\mathrm{E} & \rightarrow & \mathrm{CO(a)} \nonumber\\
\mathrm{O}_2\mathrm{(g)} + 2\mathrm{E} & \rightarrow & 2\mathrm{O(a)} \\
\mathrm{CO(a)} + \mathrm{O(a)} 
& \rightarrow & \mathrm{CO}_2\mathrm{(g)} + 2\mathrm{E} 
\;,
\nonumber
\label{eq:LH}
\end{eqnarray}
where E is an empty site on the surface, and (g) and (a) refer 
to the gas and adsorbed phase, respectively.

\begin{figure}[H]
%\centering\includegraphics[scale=1.00]{histos4.eps}
%\centering\includegraphics[scale=0.45]{histos4.eps}
\centering\input{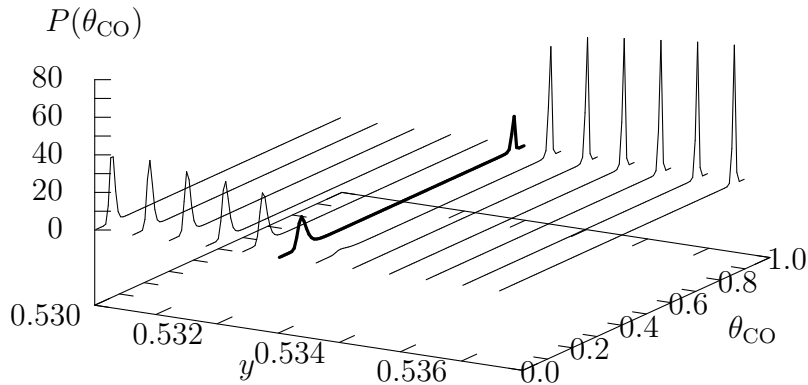}
\caption[]{Order-parameter probability distribution, 
$P(\theta_{\rm CO})$, for $k=0.02$ and $L=100$.
The distribution for the value of $y$ closest to the coexistence value, $y_2$, 
is shown with a bold line.}
\label{histos4}
\end{figure}

\vspace{15pt}
\begin{center}
SIMULATIONS AND RESULTS
\end{center}
%\vspace{5pt}
The ZGB-k model is simulated on a square lattice of linear size $L$ that 
represents the catalytic surface. The kinetic 
Monte Carlo simulation generates a 
sequence of trials: CO or O$_2$ adsorption with probability $1-k$, 
and CO desorption with probability $k$. For the adsorption, a CO or O$_2$ 
molecule is selected with probability $y$ or $1-y$, respectively.  
We calculate the  coverages $\theta_{\rm CO}$ and
$\theta_{\rm O}$,  defined as the fraction of surface sites occupied by 
CO and O, respectively, and 
the rate of production of CO$_{2}$, $R_{{\rm CO}_2}$. In Fig.~\ref{histos4}  
we show $P(\theta_{\rm CO})$, the
probability distribution for $\theta_{\rm CO}$ vs $y$, where it is clearly 
seen that at a particular value of $y$,
$y_2(k)$, a low $\theta_{\rm CO}$ and a high $\theta_{\rm O}$ phase coexist.

A finite-size-scaling analysis of the statistical fluctuations 
of the CO coverage gives strong evidence that below the critical value of  
$k$ the model exhibits a first-order, nonequilibrium phase transition 
between low and high CO coverage phases with the same characteristics as a 
first-order equilibrium phase transition [4,5,7].

We also measured the metastable lifetimes associated with the transition from 
the low CO coverage phase to the high CO coverage phase and vice versa. 
The system-size dependence of the decay times 
strongly suggests that the system follows a decay mechanism very similar to 
the one described by the
classic Kolmogorov-Johnson-Mehl-Avrami theory of phase transformation by 
nucleation and growth near
a first-order equilibrium phase transition, with well-defined single-droplet 
and multidroplet regimes [8,9]. In this system, the desorption parameter and 
the distance to the coexistence curve play
the roles of the temperature and the supersaturation 
or overpotential, respectively. Near the 
coexistence curve the decay times are inversely
proportional to $1/L²$, and the decay mechanism consists of the nucleation 
and growth of a single droplet of the stable phase. Far from the coexistence 
curve, the decay times are independent of the system size, and the decay 
proceeds by random nucleation of many droplets of the stable phase [5].

\begin{figure}
\centering\includegraphics[scale=.65, clip]{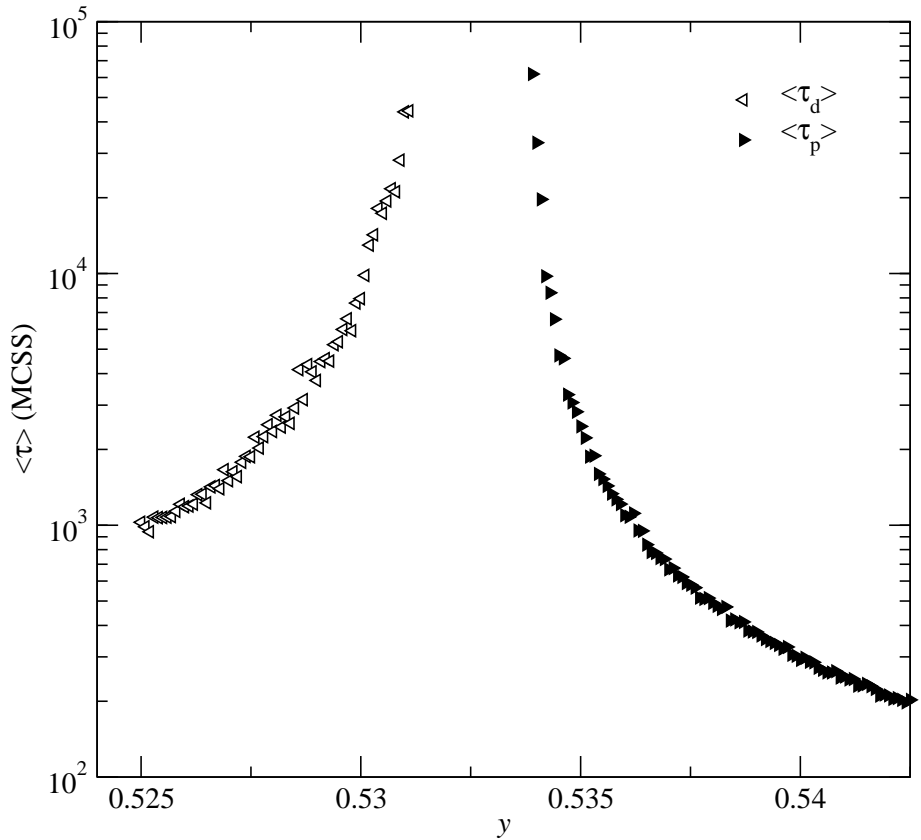}
\caption[]{Decay times
as functions of $y$ when the system evolves toward the low CO
coverage region $\langle \tau_d \rangle$, and when it evolves toward the 
high coverage region $\langle \tau_d \rangle$, shown
for $k=0.02$ and $L=100$. Here and elsewhere in this paper,
time is measured in Monte Carlo steps per site (MCSS).}
\label{decay}
\end{figure}
\begin{figure}
\centering\includegraphics[scale=0.65]{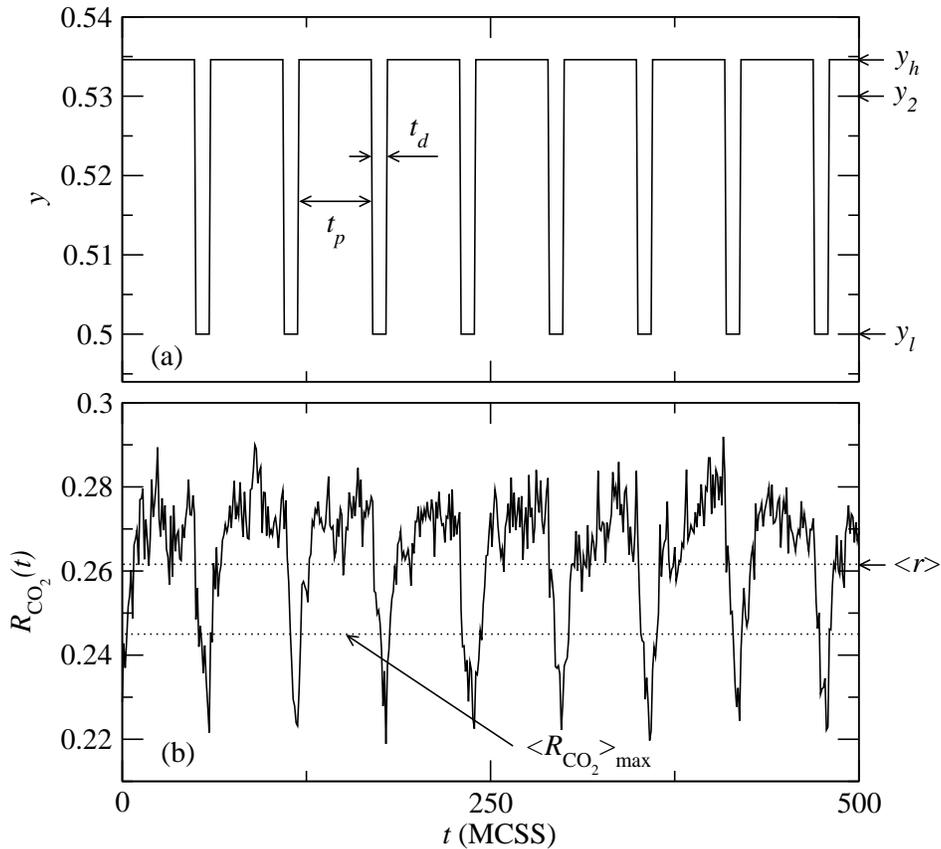}
\caption{(a) Applied periodic pressure of CO, $y(t)$, 
that takes the values
$y_l=0.5$ and $y_h=0.5346$ during the time intervals $t_d=10$ and
$t_p=50$, respectively.
(b) Response of the production rate to the
applied pressure given in (a) for $L=100$ with $k=0.02$.
The dotted line marked $\langle r \rangle $ 
indicates the long-time average of the 
period-averaged CO$_2$ production rate $r$, 
while the dotted line marked $\langle R_{\rm CO_2} \rangle_{\rm max}$ marks 
the maximum average CO$_2$ production rate for {\it constant\/} $y=y_2(k)$.
%Time is measured in units of Monte Carlo steps per site (MCSS).
}
\label{serie_osc3}
\end{figure}

We found that the lifetimes strongly depend on $k$ and on the direction of 
the process; the mean decontamination time $\langle \tau_d \rangle$ 
(from high to low CO coverage) 
being different from the mean poisoning time $\langle \tau_p \rangle$ (from
high to low CO coverage). At comparable distances from 
the coexistence curve, $\langle \tau_d \rangle \gg \langle \tau_p \rangle$, 
as seen Fig.~\ref{decay}.
Since several experiments indicate that it is possible to increase the 
efficiency of a catalytic process by subjecting the system to periodic 
forcing [10],
we decided to exploit the asymmetry between the decay times by subjecting 
the system to a periodic
variation of the external pressure with periods related to the decay times 
in each direction. We therefore select 
a square-wave periodic variation of the CO pressure, $y(t)$, that 
in a period $T=t_{d}+
t_{p}$ takes the values, $y_l$ (located below the transition pressure) 
during the time interval $t_{d}$ and $y_h$ (located above the transition 
pressure) during the time interval $t_{p}$, as can 
be seen in Fig.~\ref{serie_osc3}(a). 

We found that the times that the driving force spends in the low and high 
coverage regions, $t_d$ and $t_p$, respectively, can be tuned for each set 
of $y_l$ and $y_h$ to increase the productivity of the system. 
In Fig.~\ref{serie_osc3}(b) it is seen that the CO$_2$ production rate 
exhibits an oscillatory behavior in response to the periodic pressure shown in 
Fig.~\ref{serie_osc3}(a). The period-averaged value
of the CO$_2$ production rate $R_{\rm CO_2}$, defined as
\begin{equation}
r = \frac{1}{T}\oint R_{{\rm CO}_2}(t) \mathrm{d}t,
\label{order_parameter}
\end{equation}
plays the role of the  dynamic order parameter [6]. For the parameters 
selected in Fig.~\ref{serie_osc3}, the long-time average of $r$, 
$\langle r \rangle$ takes
a value that is about 7\% higher than the maximum average CO$_2$ 
production rate for constant $y$. It is likely that more careful 
tuning of the parameters could further improve the degree 
of enhancement.  

\begin{figure}
\centering\includegraphics[clip, scale=0.65]{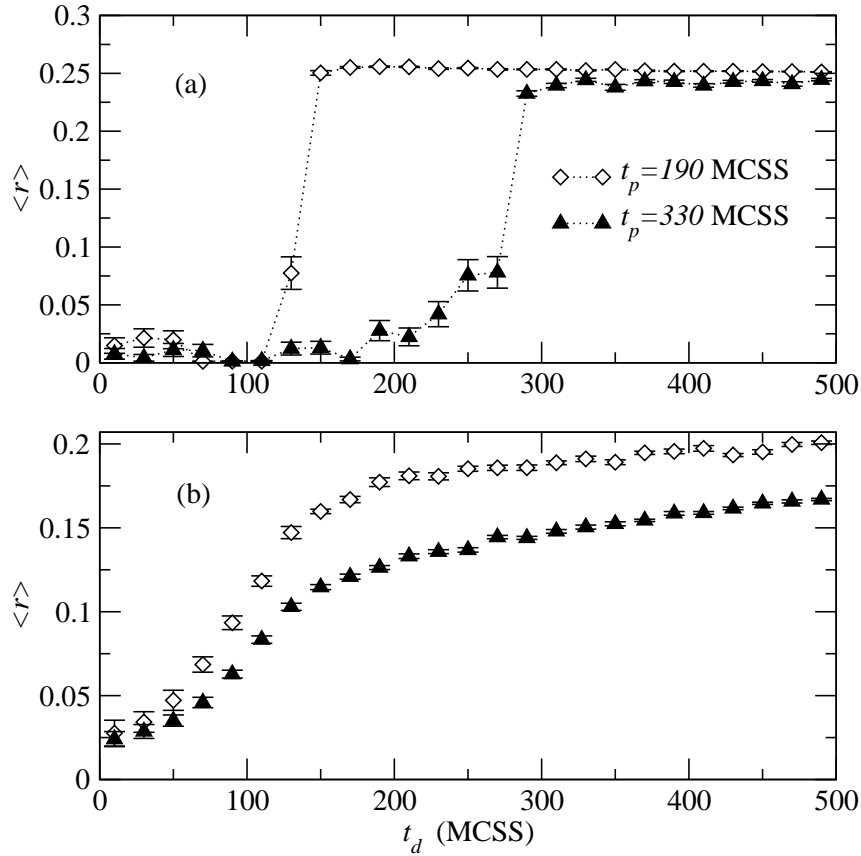}
\caption{Long-time average of the period-averaged rate of CO$_2$ production, 
$\langle r\rangle$, shown vs $t_d$ for two values of $t_p$ and $L=100$;
(a) with $y_l=0.52$, $y_h=0.535$, 
and $k=0.01$, and (b) with $y_l=0.52$, $y_h=0.553$, and $k=0.04$.
Only for the lower value of $k$ 
does the system clearly present two dynamic phases: 
one with $\langle r\rangle\approx 0$ and the other with $\langle r\rangle>0$.}
\label{r_vs_td3}
\end{figure}

We also found that, for sufficiently low values of $k$,  the driven system 
undergoes a dynamic phase transition between a dynamic phase of high CO$_2$ 
production,  $r > 0$, and a nonproductive one, $r\approx 0$, as can be seen 
in Fig.~\ref{r_vs_td3}. The distinction between these phases disappears for 
desorption rates above the critical value [6].

\begin{figure}
\centering\includegraphics[clip, scale=0.75]{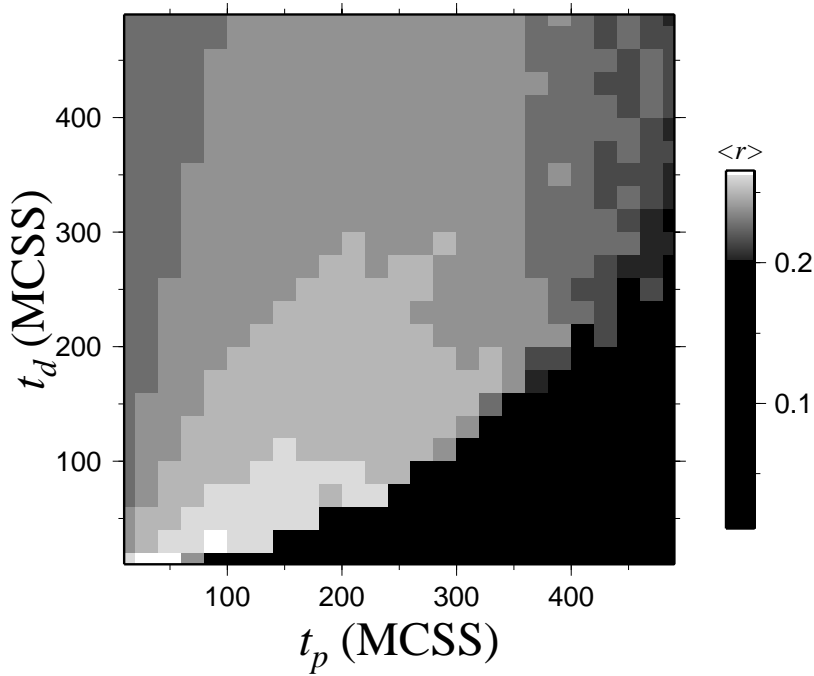}
\caption{
Long-time average $\langle r \rangle$ of the period-averaged CO$_2$ 
production rate $r$, shown as a density plot vs $t_p$ and $t_d$ for
$y_l=0.51$, $y_h=0.535$, $k=0.01$, and $L=100$.
}
\label{fig:image}
\end{figure}

A detailed 
finite-size scaling analysis indicates that for small values of $k$, the 
measure of the fluctuations of the order parameter, 
\begin{equation}
X_L = L^2 [\langle r^2 \rangle - \langle r \rangle ^2]
\;,
\label{X}
\end{equation}
diverges as 
a power law with the system size, $X_{L}^\mathrm{max} \approx L^{\gamma/\nu}$ 
with exponent $\gamma/\nu =1.77\pm 0.02$, while moments of
the order parameter at the transition point decay as 
$\langle r^n \rangle_L \approx L^{-n\beta/\nu}$ with 
$\beta/\nu=0.12\pm0.04$. These exponent 
ratios, together with general symmetry arguments, give reasonable evidence 
that this far-from equilibrium phase transition 
belongs to the same universality class as the equilibrium Ising model [6]. 
The long-time average production rate $\langle r \rangle$ is shown in a 
density plot vs $t_p$ and $t_d$ in Fig.~\ref{fig:image}. The line of 
critical points appears as the sharp boundary of the black, low-production 
region in the lower right-hand part of the figure. 
The region of maximum average 
production is seen to lie very close to the critical line on its 
high-production side. 
We believe this observation offers a clue to understanding
the physical reason for the enhancement. Most likely, the long-range critical 
correlations associated with the critical cluster that develops for parameter
values near the critical line produce a high density of sites where 
CO molecules would be situated next to O atoms, in positions conducive to 
rapid oxidation and desorption of the produced CO$_2$. 

\vspace{15pt}
\begin{center}
CONCLUSION
\end{center}
%\vspace{5pt}
In this paper we have summarized results of large-scale kinetic Monte Carlo 
simulations of the ZGB-k model of heterogeneously catalyzed 
CO oxidation with CO desorption. The results of the simulations were analyzed 
using finite-size scaling methods. 
We found that there are strong similarities between the dynamics of 
metastable decay in this far-from equilibrium, non-Hamiltonian system and the 
well-known behavior of Hamiltonian systems. From a theoretical point of 
view, these similarities could lead to significant advancement in 
our understanding of the 
dynamics of far-from-equilibrium systems. From a practical point of view, 
our results can be exploited 
to develop novel ways to increase the efficiency of catalytic reactions. 

\clearpage
%\vspace{15pt}
\begin{center}
ACKNOWLEDGMENTS
\end{center}
\noindent
This work was supported in part by U.S.\ National Science 
Foundation Grant No.\ DMR-0240078 at Florida State University and 
Grant No.\ DMS-0244419 at The University of Michigan. \\

\begin{center}
REFERENCES
\end{center}

%\begin{thebibliography}{99}

%\bibitem{general1} 
\noindent
1. H. ~J.\ Jensen, in {\it Self-Organized Criticality:
Emergent Complex Behavior in Physical and Biological
Systems}, Cambridge University Press, Cambridge (1998); 
J.~Marro and R.~Dickman, in
{\it Non-equilibrium Phase Transitions in Lattice Models},
Cambridge University Press, Cambridge (1999)\\

%\bibitem{surface}
\noindent
2. K.~Christmann, {\it Introduction to Surface
Physical Chemistry}, Steinkopff Verlag, Darmstadt (1991); V.~P.\
Zhdanov and B.~Kazemo, { Surf.\ Sci.\ Rep.} {\bf 20}, 111 (1994);
G.~C.\ Bond {\it Catalysis: Principles and
Applications}, Clarendon, Oxford (1987).\\

%\bibitem{ziff86}
\noindent
3. R.~M.\ Ziff, E.~Gulari, and Y.~Barshad, Phys.\ Rev.\ Lett.\
\textbf{56}, 2553 (1986).\\

%\bibitem{tome93}
\noindent
4. T.~Tom\'e and R.~Dickman, Phys.\ Rev.\ E
\textbf{47}, 948 (1993).\\

%\bibitem{machado1}
\noindent
5. E.~Machado, G.~M.\ Buend\y a, and P.~A.\ Rikvold, Phys.\ Rev.\ E
\textbf{71}, 031603 (2005).\\

%\bibitem{machado2}
\noindent
6.  E.~Machado, G.~M.\ Buend\y a, P.~A. Rikvold, and R.~M.\ Ziff, 
Phys.\ Rev.\ E \textbf{71}, 016120 (2005).\\

%\bibitem{ziff2}
\noindent
7. R.~M.\ Ziff and B.~J.\ Brosilow, Phys.\ Rev.\ A
\textbf{46}, 4630 (1992).\\

%\bibitem{kJMA}
\noindent
8. A.~N.\ Kolmogorov, Bull.\ Acad.\ Sci.\ USSR, Phys.\ Ser.\
{\bf 1}, 335 (1937);
W.~A.\ Johnson and R.~F.\ Mehl,
Trans.\ Am.\ Inst.\ Mining Metall.\ Eng.\ {\bf 135}, 416 (1939);
M.~Avrami, J.\ Chem.\ Phys.\ {\bf 7}, 1103 (1939);
{\bf 8}, 212 (1940); {\bf 9}, 177 (1941).\\

\noindent
9. P.~A.\ Rikvold, H.~Tomita, S.~Miyashita, and S.~W.\ Sides,
Phys.\ Rev.\ E {\bf 49}, 5080 (1994).\\ 

%\bibitem{imbhil95
\noindent
10. R.~Imbhil and G.~Ertl,  Chem.\ Rev.\
{\bf 95}, 697 (1995);
M.~Ehsasi, M.~Matloch, O.~Frank, J.~H.\ Block, K.~Christmann,
F.~S.\ Rys, and W.~Hirschwald,  J.\ Chem.\ Phys.\
{\bf 91}, 4949 (1989).\\

%\end{thebibliography}

\end{document}